\documentstyle[11pt,psfig,twoside]{article}

\begin{document}

\newcommand{\dd}{\,{\rm d}}
\newcommand{\ie}{{\it i.e.},\,}
\newcommand{\etal}{{\it et al.\ }}
\newcommand{\eg}{{\it e.g.},\,}
\newcommand{\cf}{{\it cf.\ }}
\newcommand{\vs}{{\it vs.\ }}
\newcommand{\zdot}{\makebox[0pt][l]{.}}
\newcommand{\up}[1]{\ifmmode^{\rm #1}\else$^{\rm #1}$\fi}
\newcommand{\dn}[1]{\ifmmode_{\rm #1}\else$_{\rm #1}$\fi}
\newcommand{\upd}{\up{d}}
\newcommand{\uph}{\up{h}}
\newcommand{\upm}{\up{m}}
\newcommand{\ups}{\up{s}}
\newcommand{\arcd}{\ifmmode^{\circ}\else$^{\circ}$\fi}
\newcommand{\arcm}{\ifmmode{'}\else$'$\fi}
\newcommand{\arcs}{\ifmmode{''}\else$''$\fi}
\newcommand{\MS}{{\rm M}\ifmmode_{\odot}\else$_{\odot}$\fi}
\newcommand{\RS}{{\rm R}\ifmmode_{\odot}\else$_{\odot}$\fi}
\newcommand{\LS}{{\rm L}\ifmmode_{\odot}\else$_{\odot}$\fi}

\newcommand{\Abstract}[2]{{\footnotesize\begin{center}ABSTRACT\end{center}
\vspace{1mm}\par#1\par
\noindent
{~}{\it #2}}}

\newcommand{\TabCap}[2]{\begin{center}\parbox[t]{#1}{\begin{center}
  \small {\spaceskip 2pt plus 1pt minus 1pt T a b l e}
  \refstepcounter{table}\thetable \\[2mm]
  \footnotesize #2 \end{center}}\end{center}}

\newcommand{\TableSep}[2]{\begin{table}[p]\vspace{#1}
\TabCap{#2}\end{table}}

\newcommand{\FigCap}[1]{\footnotesize\par\noindent Fig.\  %
  \refstepcounter{figure}\thefigure. #1\par}

\newcommand{\TableFont}{\footnotesize}
\newcommand{\TableFontIt}{\ttit}
\newcommand{\SetTableFont}[1]{\renewcommand{\TableFont}{#1}}

\newcommand{\MakeTable}[4]{\begin{table}[htb]\TabCap{#2}{#3}
  \begin{center} \TableFont \begin{tabular}{#1} #4 
  \end{tabular}\end{center}\end{table}}

\newcommand{\MakeTableSep}[4]{\begin{table}[p]\TabCap{#2}{#3}
  \begin{center} \TableFont \begin{tabular}{#1} #4 
  \end{tabular}\end{center}\end{table}}

\newenvironment{references}%
{
\footnotesize \frenchspacing
\renewcommand{\thesection}{}
\renewcommand{\in}{{\rm in }}
\renewcommand{\AA}{Astron.\ Astrophys.}
\newcommand{\AAS}{Astron.~Astrophys.~Suppl.~Ser.}
\newcommand{\ApJ}{Astrophys.\ J.}
\newcommand{\ApJS}{Astrophys.\ J.~Suppl.~Ser.}
\newcommand{\ApJL}{Astrophys.\ J.~Letters}
\newcommand{\AJ}{Astron.\ J.}
\newcommand{\IBVS}{IBVS}
\newcommand{\PASP}{P.A.S.P.}
\newcommand{\Acta}{Acta Astron.}
\newcommand{\MNRAS}{MNRAS}
\renewcommand{\and}{{\rm and }}
\section{{\rm REFERENCES}}
\sloppy \hyphenpenalty10000
\begin{list}{}{\leftmargin1cm\listparindent-1cm
\itemindent\listparindent\parsep0pt\itemsep0pt}}%
{\end{list}\vspace{2mm}}

\def\TYLDA{~}
\newlength{\DW}
\settowidth{\DW}{0}
\newcommand{\dw}{\hspace{\DW}}

\newcommand{\refitem}[5]{\item[]{#1} #2%
\def\REFARG{#3}\ifx\REFARG\TYLDA\else, {\it#3}\fi
\def\REFARG{#4}\ifx\REFARG\TYLDA\else, {\bf#4}\fi
\def\REFARG{#5}\ifx\REFARG\TYLDA\else, {#5}\fi.}

\newcommand{\Section}[1]{\section{#1}}
\newcommand{\Subsection}[1]{\subsection{#1}}
\newcommand{\Acknow}[1]{\par\vspace{5mm}{\bf Acknowledgements.} #1}
\pagestyle{myheadings}

\def\thefootnote{\fnsymbol{footnote}}
\begin{center}

{\Large\bf The Optical Gravitational Lensing Experiment.\\
\vskip3pt
Age of Star Clusters from the SMC\footnote{Based on 
observations obtained with the 1.3~m Warsaw telescope at the Las
Campanas Observatory of the Carnegie Institution of Washington.}}

\vskip 1cm

{G.~~P~i~e~t~r~z~y~\'n~s~k~i,~~ and~~A.~~U~d~a~l~s~k~i}
\vskip5mm
{Warsaw University Observatory, Al.~Ujazdowskie~4, 00-478~Warszawa, Poland\\
e-mail: (pietrzyn,udalski)@sirius.astrouw.edu.pl}

\end{center}

\Abstract{We present determination of age of clusters from 2.4 square degree 
region of the SMC bar. The photometric data were taken from the {\it BVI} maps 
of the SMC and catalog of clusters in this galaxy obtained during the OGLE-II 
microlensing survey. 

For 93 well populated SMC clusters their age is derived with the standard 
procedure of isochrone fitting. The distribution of age of cluster from the 
SMC is presented. It indicates either non-uniform process of cluster formation 
or very effective disruption of clusters.} {~}

\Section{Introduction}
Studies of star clusters provide important information about parent galaxy 
and processes connected with their formation and evolution. The rich system of 
clusters from the SMC is especially well suited for such investigations. In 
particular, based on the age distribution of clusters one can look into the 
cluster formation history and obtain information on processes of cluster 
formation and disruption. 

Unfortunately the SMC was neglected photometrically for years. Until recently 
only a few papers presented precise photometric data obtained with modern 
observational techniques for clusters from this galaxy. A few old clusters 
from the SMC were studied using the HST telescope (Mighell \etal 1998). 
Analysis of other SMC clusters can be found very sporadically in the 
literature. As a result precise age of the SMC clusters based on good 
photometric data has been derived for very limited group of objects. 

Information about the age of clusters from the SMC is based mostly on 
indirect methods like integrated photometry or calibration of brightness of 
the brightest star. However, the method of deriving the age from integrated 
photometric data may be affected by several sources of uncertainty (Girardi 
\etal 1995). 

In this paper we present new, homogeneous determination of age of 93 clusters 
from the center of the SMC. After presentation of the observational data, the 
procedure of deriving the mean reddening and age is described followed by 
discussion of distribution of age of the SMC clusters. 

\Section{Observational Data}
The observational data used in this paper were collected during the OGLE-II 
microlensing search with the 1.3-m Warsaw telescope located at the Las 
Campanas Observatory, Chile, which is operated by the Carnegie Institution of 
Washington. Eleven slightly overlapping fields located in the center of the 
SMC were observed. About 2.2 million stars were monitored since June 1997. The 
photometric data were calibrated to the standard {\it BVI} system based on a 
few hundred observations of standard stars from Landolt (1992) fields. The 
typical uncertainty of photometric calibration was about 0.01~mag. Astrometric 
position of every star was also derived with typical accuracy of 0.1~arcsec. 
Details on the OGLE-II project can be found in Udalski, Kubiak and 
Szyma{\'n}ski (1997). The comprehensive description of photometric and 
astrometric data is given by Udalski \etal (1998b). Using these data the 
catalog of star clusters was constructed by Pietrzy{\'n}ski \etal (1998). 
Altogether 238 objects were found. 72 of them are newly discovered clusters. 

The catalog of clusters in the SMC is very well suited for studies of 
properties of population of clusters from the SMC. In this paper we present 
age determination of large sample of objects from this catalog. 

\Section{Age of the SMC Clusters}
\Subsection{Interstellar Reddening}
Recently Paczy{\'n}ski and Stanek (1998) proposed a new method of distance 
determination based on the mean {\it I}-band brightness of the red clump 
stars. Red clump stars seem to form a very homogeneous group of objects. Their 
mean {\it I}-band magnitude does not depend on age for stars younger than 
10~Gyr, and weakly depends on metallicity (Udalski 1998a,b). Large number of 
these stars allows to derive their mean {\it I}-band magnitude with high 
precision. Recent calibration based on Hipparcos measurements of a few hundred 
nearby red clump stars makes these objects the best calibrated standard 
candles. Red clump stars were used for modeling the galactic bar (Stanek \etal 
1997) and determining the distances to the Magellanic Clouds (Udalski \etal 
1998a). They can also be used as a reference for interstellar extinction 
determination. For example, Stanek (1996) presented the extinction map of the 
Baade's Window in the Galactic bulge determined with this method. 

\begin{figure}[htb]
\psfig{figure=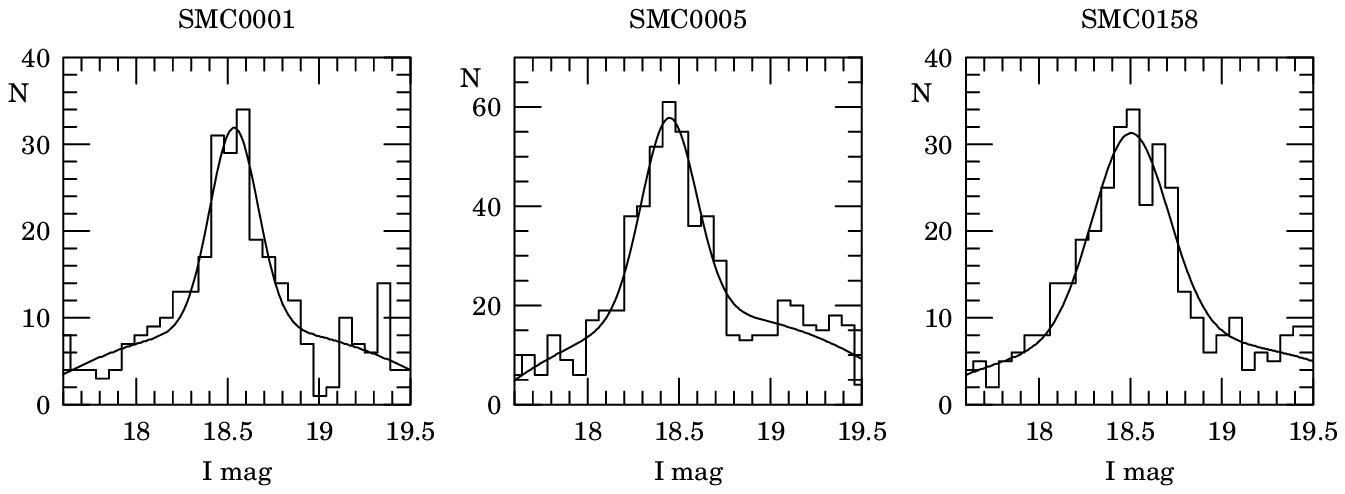,bbllx=70pt,bblly=560pt,bburx=470pt,bbury=720pt,width=12.5cm,clip=}
\FigCap{Histograms of brightness of red clump stars with fitted function given 
by Eq.~(1) for regions located around three clusters.} 
\end{figure}
In order to derive the mean reddening toward a given cluster we identified red 
clump stars located in the cluster neighborhood excluding cluster stars. Their 
mean {\it I}-band magnitude was obtained in a similar manner as in Udalski 
\etal (1998a). All objects having {\it I}-band brightness in the range of 
${17.5<I<19.5}$~mag and ${V-I}$ color index in the range of ${0.7<V-
I<1.1}$~mag were selected and histograms of the {\it I}-band brightness of 
these stars in 0.05~mag bins were constructed. Then, we fit the following 
function: 
$$n(I)=a+b(I-I^{\rm max})+c(I-I^{\rm max})^2+\frac{N_{RC}}{\sigma_{\rm 
RC}\sqrt{2\pi}}\exp\left[-\frac{(I-I^{\rm
max})^2}{2\sigma^2_{\rm RC}}\right]\eqno(1)$$
where $N_{RC}, \sigma^2_{\rm RC}$ and $I^{\rm max}$ are the number of red 
clump stars, their standard deviation of brightness and maximum brightness, 
respectively. Fig.~1 shows exemplary histograms with fitted function given by 
Eq.~(1) obtained for regions located around three clusters. It is clearly seen 
that the well pronounced red clump allows for reliable fit and thus precise 
determination of its mean {\it I}-band brightness. Assuming that extinction 
free brightness of the red clump stars in the SMC is equal to ${I=18.34}$~mag 
(Udalski 1998b) and standard extinction curve (${A_I=1.96\cdot E(B-V)}$, 
${E(V-I)=1.28\cdot E(B-V)}$, Schlegel, Finkbeiner and Davis 1998) we derived 
extinction in the {\it I}-band as well as ${E(V-I)}$ and ${E(B-V)}$ reddenings. 
The results are given in Table~1. 
\renewcommand{\TableFont}{\scriptsize}
\MakeTableSep{|c|c|c|c|c|c|c|}{10cm}{Age of the SMC clusters}{
\hline
Name  & $\alpha_{2000}$ & $\delta_{2000}$ & Radius & $E(B-V)$ & $\log t$ &
$\sigma_{\log t}$\\
OGLE-CL- & &  & [\arcs] & & &  \\ \hline
SMC0002 &  $0\uph37\upm33\zdot\ups06$ &  $-73\arcd36\arcm42\zdot\arcs6$ &  47 &   0.06 &  8.4& 0.1\\
SMC0003 &  $0\uph37\upm42\zdot\ups24$ &  $-73\arcd54\arcm29\zdot\arcs5$ &  42 &   0.10 & $>$ 9&--\\
SMC0008 &  $0\uph40\upm30\zdot\ups54$ &  $-73\arcd24\arcm10\zdot\arcs4$ &  43 &   0.07 &  8.0 &0.1\\
SMC0009 &  $0\uph40\upm44\zdot\ups11$ &  $-73\arcd23\arcm00\zdot\arcs2$ &  36 &   0.07 &  8.0&0.1\\
SMC0011 &  $0\uph41\upm06\zdot\ups16$ &  $-73\arcd21\arcm07\zdot\arcs1$ &  36 &   0.08 &  7.9&0.1\\
SMC0012 &  $0\uph41\upm23\zdot\ups78$ &  $-72\arcd53\arcm27\zdot\arcs1$ &  61 &   0.06 &  $>$ 9&--\\
SMC0013 &  $0\uph42\upm22\zdot\ups37$ &  $-73\arcd44\arcm03\zdot\arcs1$ &  23 &   0.05 &  7.3&0.3 \\
SMC0015 &  $0\uph42\upm54\zdot\ups13$ &  $-73\arcd17\arcm37\zdot\arcs0$ &  30 &   0.10 &  8.1&0.1\\
SMC0016 &  $0\uph42\upm58\zdot\ups46$ &  $-73\arcd10\arcm07\zdot\arcs2$ &  42 &   0.08 &  8.3&0.1\\
SMC0017 &  $0\uph43\upm32\zdot\ups74$ &  $-73\arcd26\arcm25\zdot\arcs4$ &  26 &   0.10 &  7.9&0.1\\
SMC0018 &  $0\uph43\upm37\zdot\ups57$ &  $-73\arcd26\arcm37\zdot\arcs9$ &  26 &   0.10 &  7.9&0.1\\
SMC0019 &  $0\uph43\upm37\zdot\ups59$ &  $-72\arcd57\arcm30\zdot\arcs9$ &  12 &   0.08 &  8.6&0.1 \\
SMC0020 &  $0\uph43\upm37\zdot\ups89$ &  $-72\arcd58\arcm48\zdot\arcs3$ &  9  &   0.08 &  8.6&0.1 \\
SMC0025 &  $0\uph45\upm13\zdot\ups88$ &  $-73\arcd13\arcm09\zdot\arcs2$ &  15 &   0.12 &  8.0&0.1\\
SMC0032 &  $0\uph45\upm54\zdot\ups33$ &  $-73\arcd30\arcm24\zdot\arcs2$ &  30 &   0.10 &  8.0&0.1\\
SMC0033 &  $0\uph46\upm12\zdot\ups26$ &  $-73\arcd23\arcm34\zdot\arcs0$ &  18 &   0.15 &  7.2&0.2\\
SMC0038 &  $0\uph47\upm06\zdot\ups15$ &  $-73\arcd15\arcm24\zdot\arcs9$ &  21 &   0.12 &  8.1&0.2\\
SMC0039 &  $0\uph47\upm11\zdot\ups61$ &  $-73\arcd28\arcm38\zdot\arcs1$ &  49 &   0.11 &  8.0&0.1\\
SMC0043 &  $0\uph47\upm52\zdot\ups38$ &  $-73\arcd13\arcm20\zdot\arcs3$ &  22 &   0.09 &  8.5&0.1\\
SMC0045 &  $0\uph48\upm00\zdot\ups68$ &  $-73\arcd29\arcm10\zdot\arcs3$ &  35 &   0.07 &  8.4&0.1\\
SMC0047 &  $0\uph48\upm28\zdot\ups14$ &  $-72\arcd59\arcm00\zdot\arcs3$ &  36 &   0.12 &  7.8&0.1\\
SMC0049 &  $0\uph48\upm37\zdot\ups47$ &  $-73\arcd24\arcm53\zdot\arcs2$ &  38 &   0.06 &  7.0&0.2\\
SMC0050 &  $0\uph48\upm59\zdot\ups02$ &  $-73\arcd09\arcm03\zdot\arcs8$ &  14 &   0.12 &  8.3&0.1\\
SMC0054 &  $0\uph49\upm17\zdot\ups60$ &  $-73\arcd22\arcm19\zdot\arcs8$ &  27 &   0.10 &  8.0&0.1\\
SMC0058 &  $0\uph49\upm45\zdot\ups43$ &  $-72\arcd51\arcm58\zdot\arcs0$ &  36 &   0.17 &  8.3&0.1\\
SMC0059 &  $0\uph50\upm16\zdot\ups06$ &  $-73\arcd01\arcm59\zdot\arcs6$ &  25 &   0.10 &  7.8&0.1\\
SMC0060 &  $0\uph50\upm21\zdot\ups95$ &  $-73\arcd23\arcm16\zdot\arcs5$ &  36 &   0.08 &  8.4&0.1\\
SMC0061 &  $0\uph50\upm00\zdot\ups26$ &  $-73\arcd15\arcm17\zdot\arcs7$ &  21 &   0.12 &  7.4&0.2 \\
SMC0064 &  $0\uph50\upm39\zdot\ups55$ &  $-72\arcd57\arcm54\zdot\arcs8$ &  36 &   0.10 &  8.1&0.1\\
SMC0066 &  $0\uph50\upm55\zdot\ups39$ &  $-73\arcd12\arcm11\zdot\arcs0$ &  17 &   0.09 &  7.8&0.1\\
SMC0067 &  $0\uph50\upm55\zdot\ups54$ &  $-72\arcd43\arcm39\zdot\arcs7$ &  42 &   0.08 &  8.2&0.1\\
SMC0068 &  $0\uph50\upm56\zdot\ups26$ &  $-73\arcd17\arcm21\zdot\arcs1$ &  55 &   0.09 &  7.7&0.2\\
SMC0069 &  $0\uph51\upm14\zdot\ups13$ &  $-73\arcd09\arcm41\zdot\arcs5$ &  36 &   0.08 &  7.6&0.1\\
SMC0070 &  $0\uph51\upm26\zdot\ups15$ &  $-73\arcd16\arcm59\zdot\arcs8$ &  14 &   0.09 &  7.8&0.1\\
SMC0071 &  $0\uph51\upm31\zdot\ups78$ &  $-73\arcd00\arcm38\zdot\arcs3$ &  32 &   0.07 & 7.5&0.1\\
SMC0072 &  $0\uph51\upm41\zdot\ups69$ &  $-73\arcd13\arcm46\zdot\arcs8$ &  28 &   0.08 &  7.6&0.2\\
SMC0073 &  $0\uph51\upm44\zdot\ups03$ &  $-72\arcd50\arcm25\zdot\arcs1$ &  42 &   0.11 &  8.2&0.2\\
SMC0074 &  $0\uph51\upm52\zdot\ups91$ &  $-72\arcd57\arcm13\zdot\arcs9$ &  42 &   0.09 &  8.1&0.1\\
SMC0075 &  $0\uph51\upm54\zdot\ups32$ &  $-73\arcd05\arcm52\zdot\arcs9$ &  15 &   0.07 &  8.4&0.1\\
SMC0076 &  $0\uph52\upm12\zdot\ups47$ &  $-72\arcd31\arcm51\zdot\arcs2$ &  29 &   0.07 &  7.4&0.3\\
SMC0077 &  $0\uph52\upm13\zdot\ups34$ &  $-73\arcd00\arcm12\zdot\arcs2$ &  18 &   0.08 &  7.9&0.2\\
SMC0078 &  $0\uph52\upm16\zdot\ups56$ &  $-73\arcd01\arcm04\zdot\arcs0$ &  36 &   0.08 &  7.9&0.1\\
SMC0081 &  $0\uph52\upm33\zdot\ups65$ &  $-72\arcd40\arcm53\zdot\arcs6$ &  31 &   0.12 &  7.4&0.3\\
SMC0082 &  $0\uph52\upm42\zdot\ups12$ &  $-72\arcd55\arcm31\zdot\arcs6$ &  36 &   0.10 &  7.8&0.3\\
SMC0083 &  $0\uph52\upm44\zdot\ups27$ &  $-72\arcd58\arcm47\zdot\arcs8$ &  29 &   0.09 &  7.8&0.2\\
SMC0087 &  $0\uph52\upm48\zdot\ups99$ &  $-73\arcd24\arcm43\zdot\arcs3$ &  22 &   0.10 &  8.7&0.1\\
SMC0089 &  $0\uph53\upm05\zdot\ups28$ &  $-72\arcd37\arcm27\zdot\arcs8$ &  118 &   0.09 &  7.3&0.4\\
\hline
}

\setcounter{table}{0}
\MakeTableSep{|c|c|c|c|c|c|c|}{10cm}{Concluded}{
\hline
Name  & $\alpha_{2000}$ & $\delta_{2000}$ & Radius & $E(B-V)$ & $\log t$ &
$\sigma_{\log t}$\\
OGLE-CL- & &  & [\arcs] & & &  \\ \hline
SMC0090 &  $0\uph53\upm05\zdot\ups59$ &  $-73\arcd22\arcm49\zdot\arcs4$ &  38 &   0.11 &  8.5 & 0.1\\
SMC0092 &  $0\uph53\upm17\zdot\ups90$ &  $-72\arcd45\arcm59\zdot\arcs5$ &  34 &   0.12 &  7.4&0.1\\
SMC0098 &  $0\uph54\upm46\zdot\ups73$ &  $-73\arcd13\arcm24\zdot\arcs5$ &  36 &   0.07 &  8.0&0.1\\
SMC0099 &  $0\uph54\upm48\zdot\ups24$ &  $-72\arcd27\arcm57\zdot\arcs8$ &  36 &   0.08 &  7.6&0.2\\
SMC0104 &  $0\uph55\upm32\zdot\ups98$ &  $-72\arcd49\arcm58\zdot\arcs1$ &  37 &   0.12 &  8.6&0.1\\
SMC0105 &  $0\uph55\upm42\zdot\ups99$ &  $-72\arcd52\arcm48\zdot\arcs4$ &  44 &   0.11 &  8.0&0.1\\
SMC0107 &  $0\uph56\upm18\zdot\ups68$ &  $-72\arcd27\arcm50\zdot\arcs4$ &  57 &   0.10 &  7.5&0.1\\
SMC0109 &  $0\uph57\upm29\zdot\ups80$ &  $-72\arcd15\arcm51\zdot\arcs9$ &  24 &   0.04 &  7.7&0.1\\
SMC0112 &  $0\uph57\upm57\zdot\ups14$ &  $-72\arcd26\arcm42\zdot\arcs0$ &  29 &   0.11 &  7.5&0.3\\
SMC0115 &  $0\uph58\upm33\zdot\ups64$ &  $-72\arcd16\arcm51\zdot\arcs6$ &  15 &   0.07 &  7.3&0.3\\
SMC0117 &  $0\uph59\upm13\zdot\ups86$ &  $-72\arcd36\arcm29\zdot\arcs3$ &  45 &   0.13 &  8.3&0.1\\
SMC0118 &  $0\uph59\upm48\zdot\ups03$ &  $-72\arcd20\arcm02\zdot\arcs5$ &  41 &   0.08 &  8.2&0.1\\
SMC0120 &  $1\uph00\upm01\zdot\ups33$ &  $-72\arcd22\arcm08\zdot\arcs7$ &  27 &   0.07 &  7.7&0.2\\
SMC0121 &  $1\uph00\upm13\zdot\ups03$ &  $-72\arcd27\arcm43\zdot\arcs8$ &  30 &   0.08 &  7.9&0.1\\
SMC0122 &  $1\uph00\upm26\zdot\ups77$ &  $-73\arcd05\arcm11\zdot\arcs6$ &  36 &   0.06 &  8.3&0.1\\
SMC0124 &  $1\uph00\upm34\zdot\ups41$ &  $-72\arcd21\arcm55\zdot\arcs8$ &  34 &   0.06 &  7.6&0.1\\
SMC0126 &  $1\uph01\upm02\zdot\ups01$ &  $-72\arcd45\arcm05\zdot\arcs2$ &  40 &   0.07 &  8.0&0.1\\
SMC0128 &  $1\uph01\upm37\zdot\ups15$ &  $-72\arcd24\arcm24\zdot\arcs7$ &  36 &   0.09 &  7.1&0.3\\
SMC0129 &  $1\uph01\upm45\zdot\ups08$ &  $-72\arcd33\arcm51\zdot\arcs8$ &  29 &   0.09 &  7.3&0.1\\
SMC0134 &  $1\uph03\upm11\zdot\ups52$ &  $-72\arcd16\arcm21\zdot\arcs0$ &  28 &   0.05 &  7.8&0.1\\
SMC0137 &  $1\uph03\upm22\zdot\ups67$ &  $-72\arcd39\arcm05\zdot\arcs6$ &  36 &   0.06 &  7.6&0.2\\
SMC0138 &  $1\uph03\upm53\zdot\ups02$ &  $-72\arcd06\arcm10\zdot\arcs5$ &  18 &   0.04 &  7.4&0.4\\
SMC0139 &  $1\uph03\upm53\zdot\ups44$ &  $-72\arcd49\arcm34\zdot\arcs2$ &  20 &   0.07 &  7.5&0.1\\
SMC0140 &  $1\uph04\upm14\zdot\ups10$ &  $-72\arcd38\arcm49\zdot\arcs1$ &  25 &   0.09 &  7.2&0.3\\
SMC0141 &  $1\uph04\upm30\zdot\ups18$ &  $-72\arcd37\arcm09\zdot\arcs4$ &  35 &   0.09 &  8.2&0.2\\
SMC0142 &  $1\uph04\upm36\zdot\ups21$ &  $-72\arcd09\arcm38\zdot\arcs5$ &  41 &   0.06 &  7.3&0.1\\
SMC0143 &  $1\uph04\upm39\zdot\ups61$ &  $-72\arcd32\arcm59\zdot\arcs7$ &  27 &   0.09 &  8.2&0.2\\
SMC0144 &  $1\uph04\upm05\zdot\ups23$ &  $-72\arcd07\arcm14\zdot\arcs6$ &  18 &   0.05 &  7.6&0.2\\
SMC0145 &  $1\uph05\upm04\zdot\ups30$ &  $-71\arcd59\arcm24\zdot\arcs8$ &  18 &   0.07 &  7.9&0.2\\
SMC0146 &  $1\uph05\upm13\zdot\ups40$ &  $-71\arcd59\arcm41\zdot\arcs8$ &  14 &   0.06 &  7.3&0.3\\
SMC0147 &  $1\uph05\upm07\zdot\ups95$ &  $-71\arcd59\arcm45\zdot\arcs1$ &  22 &   0.06 &  7.1&0.3\\
SMC0149 &  $1\uph05\upm21\zdot\ups51$ &  $-72\arcd02\arcm34\zdot\arcs7$ &  36 &   0.08 &  8.2&0.1\\
SMC0151 &  $1\uph06\upm12\zdot\ups62$ &  $-72\arcd47\arcm38\zdot\arcs7$ &  36 &   0.10 &  8.1&0.1\\
SMC0153 &  $1\uph06\upm47\zdot\ups74$ &  $-72\arcd16\arcm24\zdot\arcs5$ &  27 &   0.04 &  7.2&0.3\\
SMC0154 &  $1\uph07\upm02\zdot\ups27$ &  $-72\arcd37\arcm18\zdot\arcs2$ &  33 &   0.12 &  8.2&0.2\\
SMC0155 &  $1\uph07\upm27\zdot\ups83$ &  $-72\arcd29\arcm35\zdot\arcs5$ &  41 &   0.09 &  7.7&0.2\\
SMC0156 &  $1\uph07\upm28\zdot\ups47$ &  $-72\arcd46\arcm09\zdot\arcs5$ &  41 &   0.09 &  8.2&0.1\\
SMC0158 &  $1\uph07\upm58\zdot\ups97$ &  $-72\arcd21\arcm19\zdot\arcs5$ &  77 &   0.08 &  $>$ 9&--\\
SMC0159 &  $1\uph08\upm19\zdot\ups45$ &  $-72\arcd53\arcm02\zdot\arcs5$ &  102 &   0.04 &  $>$ 9&--\\
SMC0160 &  $1\uph08\upm37\zdot\ups48$ &  $-72\arcd26\arcm20\zdot\arcs9$ &  20 &   0.04 &  7.6&0.3\\
SMC0177 &  $0\uph44\upm55\zdot\ups05$ &  $-73\arcd10\arcm27\zdot\arcs4$ &  9 &   0.08 &  7.9&0.2\\
SMC0187 &  $0\uph47\upm05\zdot\ups87$ &  $-73\arcd22\arcm16\zdot\arcs6$ &  14 &   0.04 &  8.2&0.1\\
SMC0194 &  $0\uph49\upm05\zdot\ups58$ &  $-73\arcd21\arcm09\zdot\arcs8$ &  12 &   0.07 &  7.9&0.1\\
SMC0195 &  $0\uph49\upm16\zdot\ups45$ &  $-73\arcd14\arcm56\zdot\arcs8$ &  28 &   0.07 &  7.3&0.2\\
SMC0197 &  $0\uph50\upm03\zdot\ups82$ &  $-73\arcd23\arcm03\zdot\arcs9$ &  24 &   0.08 &  8.4&0.1\\
SMC0198 &  $0\uph50\upm07\zdot\ups51$ &  $-73\arcd11\arcm25\zdot\arcs9$ &  22 &   0.06 &  7.9&0.1\\
SMC0200 &  $0\uph50\upm38\zdot\ups98$ &  $-72\arcd58\arcm43\zdot\arcs6$ &  11 &   0.06 &  8.0&0.1\\
SMC0210 &  $0\uph52\upm30\zdot\ups30$ &  $-73\arcd02\arcm59\zdot\arcs0$ &  21 &   0.07 &  8.2&0.2\\
SMC0229 &  $0\uph58\upm38\zdot\ups08$ &  $-72\arcd14\arcm04\zdot\arcs4$ &  20 &   0.05 &  7.9&0.1\\
SMC0230 &  $1\uph00\upm33\zdot\ups15$ &  $-72\arcd15\arcm30\zdot\arcs5$ &   9 &   0.08 &  7.5&0.3\\
\hline
}

\Subsection{Age Determination}
Clear distinction between cluster and field stars is difficult in the dense 
regions in the center of the SMC. Because the color-magnitude diagrams (CMD) 
of fields around clusters very often resemble CMDs of clusters we decided to 
perform statistical subtraction of field stars from the cluster CMD. A 
procedure described by Mateo and Hodge (1986) was adopted. 

After correction for reddening and field star subtraction we derived ages 
using standard procedure of isochrone fitting. We were able to fit reasonably 
isochrones for 93 objects from the OGLE catalog of clusters (Pietrzy{\'n}ski 
\etal 1998). The remaining clusters are too poorly populated or located in 
very dense stellar regions making reliable fit impossible. 

The isochrones were taken from the library of Bertelli \etal (1994). These 
isochrones are derived from stellar models computed with the radiative 
opacities of Iglesias \etal (1992) and cover wide range of chemical 
composition and stellar masses. 

\begin{figure}[htb]
\psfig{figure=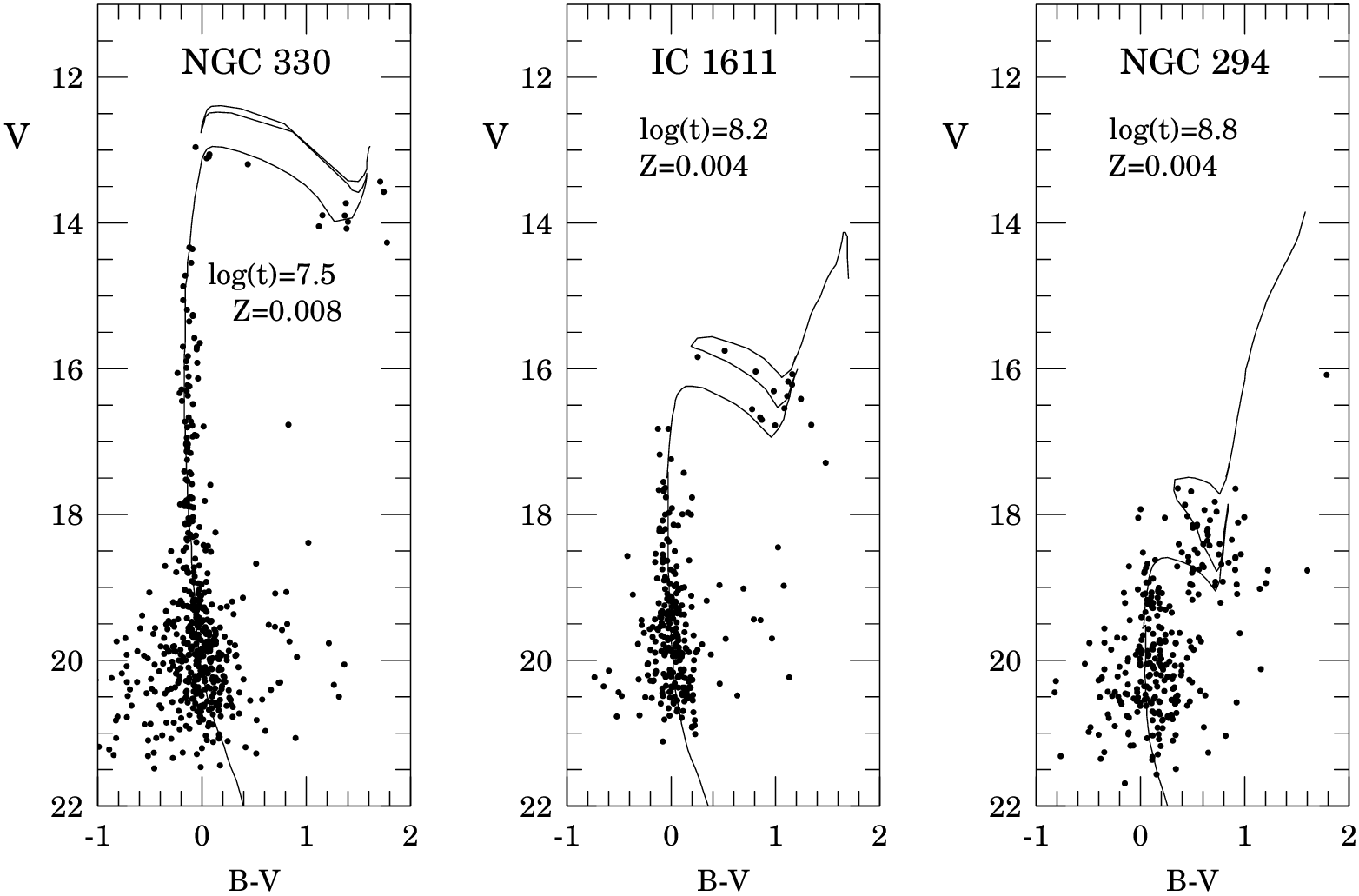,bbllx=0pt,bblly=-10pt,bburx=720pt,bbury=470pt,width=12.5cm,clip=}
\FigCap{CMDs with fitted isochrone for three clusters of different age.}
\end{figure}
The metallicity of the SMC was assumed to be ${Z=0.004}$. In the case of 
NGC~330 the models with $Z=0.008$ were used. We adopted the short distance 
scale to the SMC, namely distance modulus of 18.65~mag (Udalski 1998b). 

Results of age determination are presented in Table~1. The first column 
contains name of cluster. In the second and third columns equatorial 
coordinates are given. Cluster radius, the mean reddening toward the cluster, 
${E(B-V)}$, and determined age are presented in columns 4, 5 and 6 
respectively.

Accuracy of our procedure of age determination depends on the age and richness 
of a given cluster. For young objects as well for those possessing few stars a 
wide variety of isochrones may be fitted. The age of older clusters having 
turn-off point close to detection limit of our photometry is also less 
accurate. We estimated accuracy of age determination as half of the age 
difference between two marginally fitting isochrones selected around the best 
fit isochrone. The error of age determination is given in the last column of 
Table~1. 

Fig.~2 displays CMDs of three well populated clusters of different age: 
NGC~330, IC~1611 and NGC~294 with the best fit isochrone. 

\Subsection{Age Distribution}
Fig.~3 presents distribution of age of 93 clusters from the central regions of 
the SMC. 
 
\begin{figure}[htb]
\psfig{figure=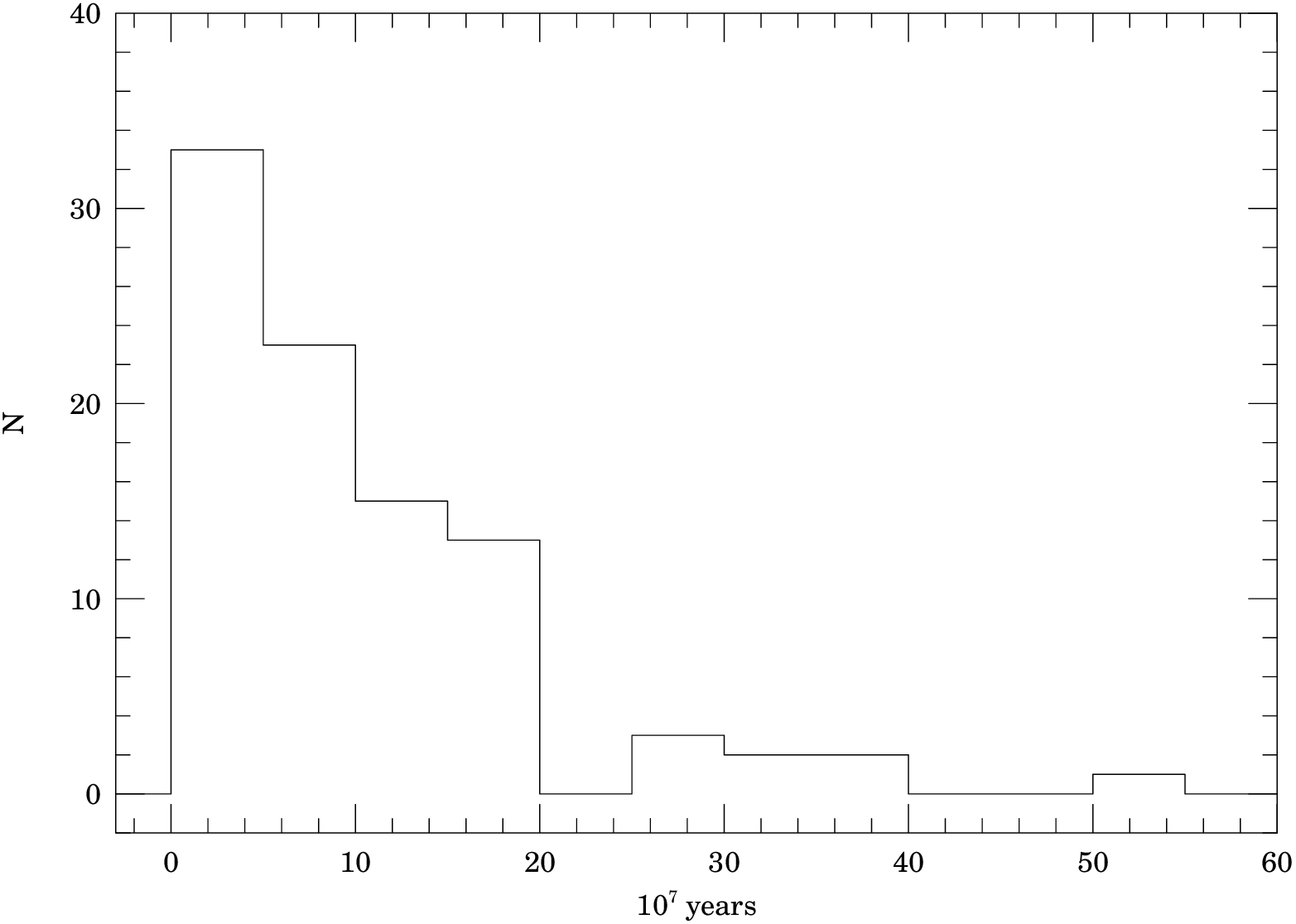,bbllx=20pt,bblly=40pt,bburx=740pt,bbury=550pt,width=12.5cm,clip=}
\FigCap{Age distribution of 93 clusters from the center of the SMC.}
\end{figure}
Fig.~3 indicates that the cluster formation rate was not uniform in the past. 
Most clusters are objects younger than $20\cdot10^{7}$ years. However, it 
should be stressed that the presented distribution does not reflect only the 
rate of formation of clusters. It is also affected by a process of dissolution 
of clusters and possible selection effects connected with detection of 
clusters and age determination. 

Our sample of the SMC clusters with determined age becomes incomplete for 
objects older than $10^{9}$ years because of limit of photometry and 
crowded field surrounding clusters. Objects with the age smaller than 
$20\cdot10^{7}$ years have the turn-off point well above the limit of 
photometric data and selection effects associated with detection and age 
determination should be negligible. Therefore, we can conclude that process of 
formation of clusters during the last ${20-30\cdot10^{7}}$ years in the 
central parts of the SMC was either very non-uniform or process of 
disintegration of clusters in this galaxy is very efficient. 

If we include to our sample two remaining oldest clusters: NGC~416 and 
NGC~419, adopt their ages from Mighell \etal (1998) and then compare the age 
distribution of clusters with that obtained for the SMC by Hodge (1987) and 
Galaxy by Wielen (1971) we may notice an evident deficiency of older 
clusters. It may reflect the incompleteness of our catalog for clusters older 
than $10^{9}$~years. On the other hand the dynamics of cluster disintegration 
in the central parts of the SMC suggests smaller number of older clusters in 
the center of the SMC than in the outer parts of this galaxy. It would be then 
important to look for older clusters in the central part of the SMC with 
deeper range photometric data than presented in this paper to confirm whether 
deficiency of older clusters in the SMC center is real. 
\vspace*{-9pt}
\Section{Summary}
\vspace*{-4pt}
We present age determination for numerous group of star clusters from the 
center of the SMC. The age of each cluster from the sample was determined 
using the standard procedure of isochrone fitting. Young clusters are the most 
frequent objects in the investigated group. The age distribution may be 
explained by non-uniform rate of cluster formation in this galaxy 
and/or strong disintegration processes. Based on these data we 
cannot, however, conclude whether the old clusters in the center of the SMC 
are less frequent than in the outer parts of this galaxy or in the Galaxy. 

\Acknow{The paper was partly supported by the Polish KBN grant 2P03D00814 to 
A.~Udalski. Partial support for the OGLE project was provided with the NSF 
grant AST-9530478 to B.~Paczy\'nski.}

\end{document}